# Investigation on the star NSVS 1925037

AMICO GIULIA[1] AND AMICO MATTIA[1]
BENNA CARLO[2], GARDIOL DANIELE[2] AND PETTITI GIUSEPPE[2]

1) IIS Curie Vittorini, Corso Allamano 130, 10095, Grugliasco (TO), Italy, TOIS03400P@istruzione.it

2) INAF-Osservatorio Astrofisico di Torino, via Osservatorio 20, I-10025 Pino Torinese (TO), Italy, giuseppe.pettiti@inaf.it

**Abstract:** an analysis of available photometric data of NSVS 1925037 and a search of a variable counterpart has been performed using astrometric and photometric data from Gaia DR2 and ASAS-SN databases.

## 1  Introduction

This report describes our investigations on star NSVS 1925037. According to the Northern Sky Variability Survey data (Wozniak, 2004), this object is of magnitude 14.469 ± 0.131 (Median ROTSE mag) and is located at R.A. 02h 39m 38.01s and DEC. +59° 04' 22.08" (J2000.0). This star was initially associated to the eclipsing binary V683 Cas (Gettel, 2006) but later analysis (Hoffman, 2009) showed that this association is incorrect. Starting from the above equatorial coordinates we tried to identify a correspondence with a known variable or a stellar object that shows variability.

## 2  Data analysis

As shown in Figure 1, the equatorial coordinates of NSVS 1925037 do not correspond to the position of any known object in the SIMBAD database. To try to identify a counterpart of this object we looked for similar objects in terms of magnitude and position.

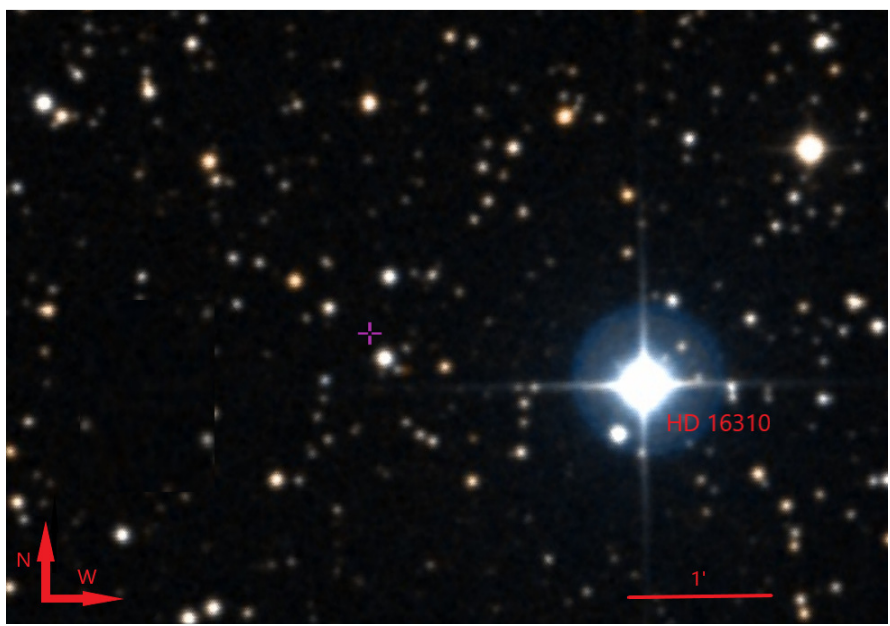

Figure 1 - Position of NSVS 1925037 (pointer)

As first step, we therefore downloaded from the Gaia DR2 database all the stars within a circle of 1' radius centered on the equatorial coordinates of NSVS 1925037, available from the Northern Sky Variability Survey. Table 1 summarizes the 57 sources present within this circle and provides their equatorial coordinates, Gaia DR2 source ID and magnitude, with error, in the G band.

| # | R.A. (J2000) | Dec. (J2000) | Source ID | G mag | e mag (±) |
|---|---|---|---|---|---|
| 1 | 02 39 37.2491673943 | +59 04 10.915927685 | 464213469952346880 | 13.9885 | 0.0063 |
| 2 | 02 39 37.1354855497 | +59 04 32.961656900 | 464213706169765376 | 20.6765 | 0.0111 |
| 3 | 02 39 36.0658053693 | +59 04 15.043879757 | 464213465651579648 | 18.2369 | 0.0020 |
| 4 | 02 39 35.6828617893 | +59 04 21.770244245 | 464213676104738816 | 19.8928 | 0.0056 |
| 5 | 02 39 40.3580420026 | +59 04 24.975451002 | 464213504308344960 | 20.0937 | 0.0057 |
| 6 | 02 39 37.4150685261 | +59 04 03.948880870 | 464213469946317312 | 19.5431 | 0.0045 |
| 7 | 02 39 35.6144120478 | +59 04 16.079451744 | 464213676111560576 | 20.7481 | 0.0106 |
| 8 | 02 39 36.5384128568 | +59 04 05.873732775 | 464213465652526464 | 17.9749 | 0.0017 |
| 9 | 02 39 40.2900450172 | +59 04 32.379431581 | 464213504312079872 | 15.5594 | 0.0004 |
| 10 | 02 39 40.4966287702 | +59 04 32.064172670 | 464213504309500928 | 20.9153 | 0.0260 |
| 11 | 02 39 35.2844770465 | +59 04 31.195308236 | 464213676107679488 | 20.9562 | 0.0219 |
| 12 | 02 39 36.9869659933 | +59 04 45.923151697 | 464213710470508544 | 14.8453 | 0.0003 |
| 13 | 02 39 34.5250813640 | +59 04 15.839548371 | 464213671810956416 | 20.2930 | 0.0065 |
| 14 | 02 39 41.5179217902 | +59 04 28.090467307 | 464213500011332224 | 18.4294 | 0.0020 |
| 15 | 02 39 40.5059710718 | +59 04 01.604126151 | 464213504312085888 | 17.0716 | 0.0160 |
| 16 | 02 39 36.0248690761 | +59 03 58.367821812 | 464213469949240832 | 20.6323 | 0.0103 |
| 17 | 02 39 34.8957795372 | +59 04 40.773258944 | 464213676107035520 | 20.4366 | 0.0083 |
| 18 | 02 39 35.6125508258 | +59 04 47.403523895 | 464213706169785984 | 18.7189 | 0.0023 |
| 19 | 02 39 35.1353432947 | +59 04 44.629034758 | 464213710466773760 | 20.2293 | 0.0069 |
| 20 | 02 39 33.9488787091 | +59 04 33.051221243 | 464213676107680512 | 20.6324 | 0.0109 |
| 21 | 02 39 40.3042874652 | +59 04 50.833676410 | 464213504306019712 | 18.7766 | 0.0033 |
| 22 | 02 39 40.5772480178 | +59 03 53.651969182 | 464213504312087168 | 18.7066 | 0.0034 |
| 23 | 02 39 33.9408397589 | +59 04 07.196288027 | 464213676110779264 | 15.8989 | 0.0006 |
| 24 | 02 39 35.5926487322 | +59 04 51.922649706 | 464213710467422976 | 20.8920 | 0.0194 |
| 25 | 02 39 40.5024129598 | +59 04 52.204667203 | 464213500011358848 | 18.2203 | 0.0017 |
| 26 | 02 39 34.6106378328 | +59 04 46.534313141 | 464213710470509440 | 16.8650 | 0.0010 |
| 27 | 02 39 37.5646392173 | +59 04 58.316295260 | 464213710467422720 | 20.5307 | 0.0098 |
| 28 | 02 39 42.1233626961 | +59 04 43.790800629 | 464213504312076672 | 14.7630 | 0.0004 |
| 29 | 02 39 42.3732637351 | +59 04 41.820118576 | 464213504306022016 | 20.9359 | 0.0218 |
| 30 | 02 39 34.2532304671 | +59 04 50.684048463 | 464213710464459648 | 19.8906 | 0.0055 |
| 31 | 02 39 38.7075132329 | +59 03 41.743470269 | 464213469946327168 | 19.9348 | 0.0057 |
| 32 | 02 39 33.1282827038 | +59 04 39.079938369 | 464213671810036864 | 18.5143 | 0.0030 |
| 33 | 02 39 36.0760712309 | +59 03 43.409715575 | 464213469952352768 | 17.7934 | 0.0014 |
| 34 | 02 39 33.1611849795 | +59 04 40.010968310 | 464213676104731008 | 20.4510 | 0.0127 |
| 35 | 02 39 40.9029753700 | +59 03 45.979440821 | 464213500011288064 | 18.2449 | 0.0019 |
| 36 | 02 39 38.8631337351 | +59 03 39.589568546 | 464213465651541632 | 18.8166 | 0.0028 |
| 37 | 02 39 43.6382771981 | +59 04 12.166204170 | 464213431291837312 | 19.4011 | 0.0034 |
| 38 | 02 39 33.1233759778 | +59 03 57.675721886 | 464213671810951808 | 20.6429 | 0.0123 |
| 39 | 02 39 32.2779232594 | +59 04 13.026000223 | 464213671810008576 | 20.5504 | 0.0095 |
| 40 | 02 39 34.0986423190 | +59 03 46.534122188 | 464213469946330624 | 20.3820 | 0.0079 |
| 41 | 02 39 43.5436784089 | +59 04 03.156372752 | 464213435589498240 | 21.0159 | 0.0173 |

| # | R.A. (J2000) | Dec. (J2000) | Source ID | G mag | e mag (±) |
|---|---|---|---|---|---|
| 42 | 02 39 33.5769515519 | +59 03 50.138119549 | 464213671809982976 | 19.6572 | 0.0049 |
| 43 | 02 39 43.1851188809 | +59 04 47.277347938 | 464219379824244608 | 20.6103 | 0.0091 |
| 44 | 02 39 35.3174396357 | +59 03 38.096341301 | 464213469952353920 | 17.3248 | 0.0009 |
| 45 | 02 39 40.2980625710 | +59 05 07.631753367 | 464213504308991872 | 20.7668 | 0.0125 |
| 46 | 02 39 40.2950435172 | +59 03 36.074817787 | 464213469952351872 | 17.8485 | 0.0013 |
| 47 | 02 39 31.5126884335 | +59 04 26.354477182 | 464213671810021248 | 19.0585 | 0.0031 |
| 48 | 02 39 37.8784874804 | +59 05 12.892292898 | 464213710470503552 | 19.4199 | 0.0057 |
| 49 | 02 39 31.5958460832 | +59 04 08.918973572 | 464213676107677568 | 21.1816 | 0.0352 |
| 50 | 02 39 34.6338274508 | +59 03 35.323387046 | 464213469952355840 | 16.8811 | 0.0008 |
| 51 | 02 39 41.3654590764 | +59 03 33.086569243 | 464213401226852224 | 19.9053 | 0.0052 |
| 52 | 02 39 44.1004113447 | +59 04 51.547682484 | 464219379827336064 | 17.2136 | 0.0010 |
| 53 | 02 39 31.6336042253 | +59 03 55.081295280 | 464213568735743744 | 19.7630 | 0.0057 |
| 54 | 02 39 45.3126246902 | +59 04 33.957653022 | 464219379827338368 | 17.2851 | 0.0009 |
| 55 | 02 39 41.1904170069 | +59 03 29.872800100 | 464213396932055552 | 18.6506 | 0.0021 |
| 56 | 02 39 37.5089585314 | +59 03 23.946882208 | 464213469946351872 | 19.9723 | 0.0059 |
| 57 | 02 39 36.0678751360 | +59 05 18.504091808 | 464213710470503296 | 18.7340 | 0.0022 |

Table 1 - Stars from Gaia DR2 database within a circle of 1' radius

Considering that the magnitude of NSVS 1925037 is available in a magnitude system slightly different from the Gaia one, we selected from Table 1 those stars with a magnitude between 12.5 and 16.5 in the G band. The selected five stars that could have a correspondence with NSVS 1925037 are listed in Table 2 and their positions with respect to NSVS 1925037 is shown in Figure 2.

| # | R.A. (J2000) | DEC (J2000) | Source ID | G mag | e mag (±) |
|---|---|---|---|---|---|
| A | 02 39 37.2491673943 | +59 04 10.915927685 | 464213469952346880 | 13.9885 | 0.0063 |
| B | 02 39 42.1233626961 | +59 04 43.790800629 | 464213504312076672 | 14.7630 | 0.0004 |
| C | 02 39 36.9869659933 | +59 04 45.923151697 | 464213710470508544 | 14.8453 | 0.0003 |
| D | 02 39 40.2900450172 | +59 04 32.379431581 | 464213504312079872 | 15.5594 | 0.0004 |
| E | 02 39 33.9408397589 | +59 04 07.196288027 | 464213676110779264 | 15.8989 | 0.0006 |

Table 2 - Selected stars that could correspond to NSVS 1925037

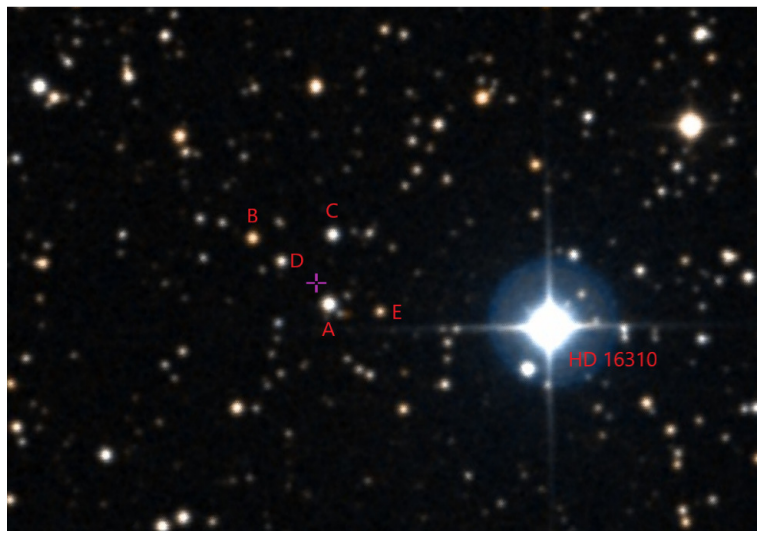

Figure 2 - Selected stars map

We then downloaded from the ASAS-SN database the photometric data and light curves of sources corresponding to the equatorial coordinates of the stars selected in Table 2. For the Gaia DR2 source ID 464213469952346880 (star A in Table 2), the ASAS-SN database contains, within 10" from the input coordinates, the star ASASSN-V J023937.20+590410.9 (or WISEJ023937.2+590410), that has already been reported to AAVSO as variable.

The ASASSN-V J023937.20+590410.9 is a star of 14.35 mag. in V, classed as an eclipsing binary. The main characteristics of this variable available from ASAS-SN (Jayasinghe, 2018b), and WISE (Chen, 2018) catalogues are summarized in Table 3. For all the other stars of Table 2, no variable counterpart is reported by ASAS-SN database.

| Source | mag. | Period (days) | Epoch (HJD) | Var. Type | Amplitude (mag) |
|---|---|---|---|---|---|
| **ASAS-SN** | 14.35 (V) | 4.4509094 | 2457613.04705 | EB | 0.34 |
| **AAVSO (WISE catalogue, Chen 2018)** | 11.611 ± 0.370 (W1) | 4.4520245 | --- | EB | --- |

Table 3 - Light curve characteristics of ASASSN-V J023937.20+590410.9

Finally, with the aim to identify or confirm a correspondence between NSVS 1925037 and one of the selected stars of Table 2, we analyzed all the photometric data and light curves, available from the Northern Sky Variability Survey and the ASAS-SN database, using the version 2.60 of the light curve and period analysis software PERANSO (Paunzen and Vanmunster, 2016).

The light curve of NSVS 1925037 available from the Northern Sky Variability Survey is shown in Figure 3. The details of the related photometric data are summarized in Table 4.

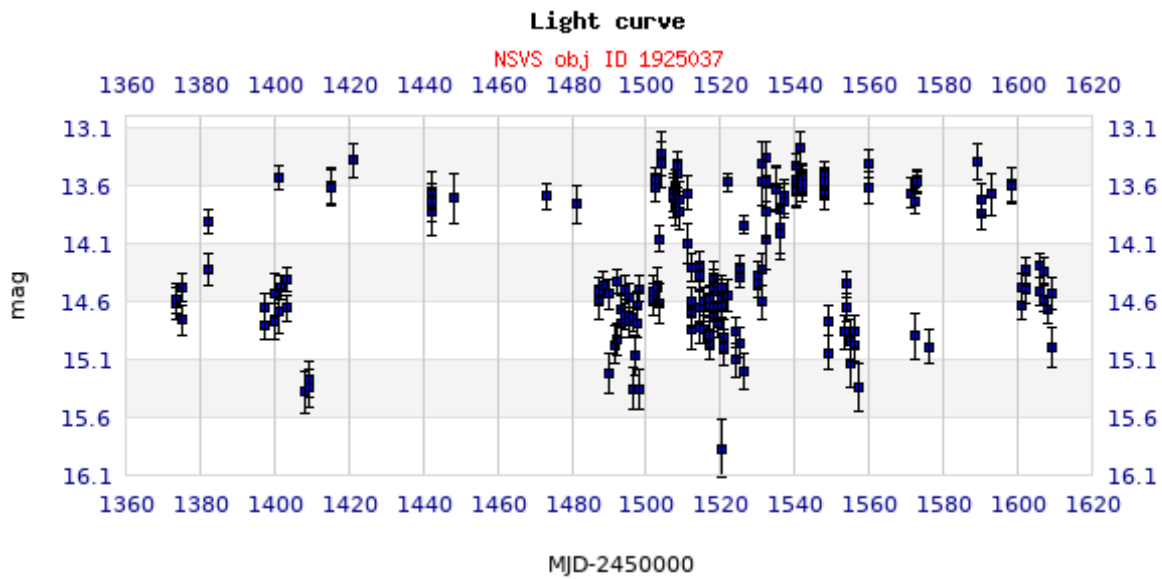

Figure 3 - Light curve of NSVS 1925037 from the Northern Sky Variability Survey

| MJD-50000 | HJD | mag (ROTSE) | err (±) | MJD-50000 | HJD | mag (ROTSE) | err (±) |
|---|---|---|---|---|---|---|---|
| 1373.357133 | 2451373.854464 | 14.617 | 0.139 | 1518.318903 | 2451518.823049 | 14.488 | 0.114 |
| 1373.358153 | 2451373.855484 | 14.577 | 0.125 | 1518.319923 | 2451518.824069 | 14.737 | 0.120 |
| 1375.349203 | 2451375.846653 | 14.487 | 0.121 | 1519.126673 | 2451519.630802 | 14.634 | 0.113 |
| 1375.350213 | 2451375.847663 | 14.757 | 0.129 | 1519.127693 | 2451519.631822 | 14.523 | 0.103 |
| 1382.349503 | 2451382.847393 | 14.328 | 0.136 | 1519.318973 | 2451519.823097 | 14.794 | 0.120 |
| 1382.350523 | 2451382.848413 | 13.911 | 0.104 | 1519.319993 | 2451519.824117 | 14.657 | 0.113 |
| 1397.336933 | 2451397.835855 | 14.803 | 0.121 | 1520.126763 | 2451520.630869 | 14.563 | 0.111 |
| 1397.337943 | 2451397.836865 | 14.650 | 0.119 | 1520.127783 | 2451520.631889 | 14.636 | 0.111 |
| 1400.247783 | 2451400.746915 | 14.773 | 0.151 | 1520.316963 | 2451520.821065 | 15.875 | 0.249 |
| 1400.248803 | 2451400.747935 | 14.539 | 0.181 | 1520.317983 | 2451520.822085 | 14.489 | 0.112 |
| 1401.287813 | 2451401.787021 | 13.539 | 0.103 | 1521.132643 | 2451521.636726 | 14.904 | 0.142 |
| 1401.288833 | 2451401.788041 | 14.689 | 0.185 | 1521.133653 | 2451521.637736 | 15.002 | 0.152 |
| 1401.439653 | 2451401.938872 | 14.480 | 0.098 | 1522.127053 | 2451522.631111 | 14.554 | 0.137 |
| 1403.335193 | 2451403.834551 | 14.419 | 0.104 | 1522.128073 | 2451522.632131 | 13.574 | 0.082 |
| 1403.336213 | 2451403.835571 | 14.661 | 0.112 | 1524.127393 | 2451524.631398 | 14.868 | 0.127 |
| 1408.326433 | 2451408.826158 | 15.380 | 0.181 | 1524.128413 | 2451524.632418 | 15.108 | 0.147 |
| 1409.370383 | 2451409.870186 | 15.276 | 0.150 | 1525.127583 | 2451525.631559 | 14.302 | 0.096 |
| 1409.371403 | 2451409.871206 | 15.347 | 0.167 | 1525.128593 | 2451525.632569 | 14.387 | 0.100 |
| 1415.313836 | 2451415.814079 | 13.610 | 0.160 | 1525.310553 | 2451525.814524 | 14.955 | 0.125 |
| 1415.369386 | 2451415.869633 | 13.612 | 0.147 | 1526.127763 | 2451526.631709 | 13.939 | 0.084 |
| 1415.369696 | 2451415.869943 | 13.622 | 0.153 | 1526.128783 | 2451526.632729 | 15.203 | 0.159 |
| 1421.238026 | 2451421.738706 | 13.387 | 0.148 | 1530.278593 | 2451530.782403 | 14.453 | 0.116 |
| 1442.114146 | 2451442.616280 | 13.833 | 0.195 | 1530.279613 | 2451530.783423 | 14.376 | 0.121 |
| 1442.169666 | 2451442.671804 | 13.651 | 0.162 | 1531.130103 | 2451531.633882 | 13.419 | 0.189 |
| 1442.169986 | 2451442.672124 | 13.748 | 0.170 | 1531.131123 | 2451531.634902 | 13.574 | 0.203 |
| 1448.107796 | 2451448.610306 | 13.707 | 0.216 | 1531.319153 | 2451531.822925 | 14.325 | 0.131 |
| 1473.107773 | 2451473.611521 | 13.691 | 0.114 | 1531.320163 | 2451531.823935 | 14.607 | 0.150 |
| 1481.078256 | 2451481.582262 | 13.763 | 0.165 | 1532.111856 | 2451532.615599 | 13.368 | 0.147 |
| 1487.093773 | 2451487.597923 | 14.603 | 0.146 | 1532.112176 | 2451532.615919 | 13.834 | 0.221 |
| 1487.250743 | 2451487.754896 | 14.508 | 0.109 | 1532.204166 | 2451532.707906 | 14.063 | 0.265 |
| 1488.248533 | 2451488.752705 | 14.457 | 0.109 | 1532.204486 | 2451532.708226 | 13.562 | 0.171 |
| 1490.132353 | 2451490.636558 | 14.539 | 0.134 | 1535.112856 | 2451535.616481 | 13.621 | 0.183 |
| 1490.133363 | 2451490.637568 | 15.221 | 0.171 | 1535.201856 | 2451535.705478 | 13.635 | 0.189 |
| 1491.361783 | 2451491.866007 | 14.974 | 0.169 | 1536.113216 | 2451536.616800 | 13.801 | 0.181 |
| 1492.317513 | 2451492.821751 | 14.927 | 0.133 | 1536.205566 | 2451536.709146 | 13.969 | 0.212 |
| 1492.318533 | 2451492.822771 | 14.435 | 0.105 | 1536.205886 | 2451536.709466 | 14.019 | 0.226 |
| 1493.317193 | 2451493.821444 | 14.790 | 0.128 | 1537.113616 | 2451537.617157 | 13.693 | 0.146 |
| 1493.318213 | 2451493.822464 | 14.675 | 0.123 | 1537.113926 | 2451537.617467 | 13.737 | 0.149 |
| 1494.358443 | 2451494.862706 | 14.504 | 0.120 | 1540.114826 | 2451540.618233 | 13.438 | 0.109 |
| 1494.359463 | 2451494.863726 | 14.719 | 0.126 | 1540.115146 | 2451540.618553 | 13.640 | 0.131 |
| 1495.318323 | 2451495.822596 | 14.567 | 0.124 | 1540.270606 | 2451540.774005 | 13.649 | 0.137 |
| 1495.319343 | 2451495.823616 | 14.779 | 0.126 | 1540.270926 | 2451540.774325 | 13.554 | 0.126 |
| 1496.315853 | 2451496.820135 | 15.358 | 0.182 | 1541.210166 | 2451541.713521 | 13.282 | 0.146 |
| 1496.316873 | 2451496.821155 | 14.738 | 0.128 | 1542.115656 | 2451542.618968 | 13.554 | 0.120 |
| 1497.130463 | 2451497.634751 | 15.071 | 0.174 | 1542.115976 | 2451542.619288 | 13.620 | 0.127 |
| 1497.313583 | 2451497.817872 | 14.639 | 0.123 | 1542.209426 | 2451542.712733 | 13.509 | 0.104 |
| 1497.314603 | 2451497.818892 | 14.786 | 0.127 | 1542.209746 | 2451542.713053 | 13.581 | 0.116 |

| MJD-50000 | HJD | mag (ROTSE) | err (±) | MJD-50000 | HJD | mag (ROTSE) | err (±) |
|---|---|---|---|---|---|---|---|
| 1498.313003 | 2451498.817299 | 15.354 | 0.170 | 1548.117476 | 2451548.620478 | 13.518 | 0.117 |
| 1498.314023 | 2451498.818319 | 14.493 | 0.114 | 1548.117786 | 2451548.620788 | 13.675 | 0.140 |
| 1502.127903 | 2451502.632211 | 14.511 | 0.129 | 1548.205676 | 2451548.708674 | 13.593 | 0.128 |
| 1502.128923 | 2451502.633231 | 14.585 | 0.139 | 1548.205996 | 2451548.708994 | 13.564 | 0.122 |
| 1502.309833 | 2451502.814141 | 13.620 | 0.125 | 1549.138033 | 2451549.640979 | 15.041 | 0.145 |
| 1502.310843 | 2451502.815151 | 13.551 | 0.097 | 1549.139053 | 2451549.641999 | 14.773 | 0.130 |
| 1503.127673 | 2451503.631981 | 14.462 | 0.158 | 1553.139603 | 2451553.642321 | 14.867 | 0.141 |
| 1503.128683 | 2451503.632991 | 14.475 | 0.159 | 1554.135013 | 2451554.637672 | 14.653 | 0.119 |
| 1503.307553 | 2451503.811860 | 14.061 | 0.116 | 1554.136023 | 2451554.638681 | 14.455 | 0.114 |
| 1503.308563 | 2451503.812870 | 14.619 | 0.169 | 1555.157183 | 2451555.659780 | 15.142 | 0.198 |
| 1504.187366 | 2451504.691672 | 13.331 | 0.186 | 1555.158203 | 2451555.660800 | 14.939 | 0.186 |
| 1504.274486 | 2451504.778792 | 13.418 | 0.194 | 1556.141233 | 2451556.643770 | 14.857 | 0.140 |
| 1507.217856 | 2451507.722150 | 13.661 | 0.162 | 1556.142243 | 2451556.644780 | 14.973 | 0.148 |
| 1507.218166 | 2451507.722460 | 13.659 | 0.164 | 1557.216833 | 2451557.719304 | 15.342 | 0.214 |
| 1507.390386 | 2451507.894679 | 13.701 | 0.171 | 1560.143493 | 2451560.645778 | 13.413 | 0.116 |
| 1508.108246 | 2451508.612534 | 13.596 | 0.132 | 1560.144503 | 2451560.646788 | 13.622 | 0.132 |
| 1508.108566 | 2451508.612854 | 13.787 | 0.152 | 1571.108503 | 2451571.610045 | 13.668 | 0.127 |
| 1508.280376 | 2451508.784663 | 13.425 | 0.117 | 1572.109123 | 2451572.610593 | 14.902 | 0.202 |
| 1508.280696 | 2451508.784983 | 13.502 | 0.123 | 1572.110143 | 2451572.611613 | 13.735 | 0.110 |
| 1509.108136 | 2451509.612416 | 13.820 | 0.158 | 1573.109783 | 2451573.611182 | 13.556 | 0.093 |
| 1509.108456 | 2451509.612736 | 13.718 | 0.144 | 1573.110803 | 2451573.612202 | 13.570 | 0.096 |
| 1511.107996 | 2451511.612256 | 14.106 | 0.168 | 1576.111683 | 2451576.612864 | 14.995 | 0.149 |
| 1511.108306 | 2451511.612566 | 13.664 | 0.142 | 1589.120053 | 2451589.620262 | 13.398 | 0.158 |
| 1512.126453 | 2451512.630701 | 14.610 | 0.123 | 1590.120673 | 2451590.620806 | 13.728 | 0.139 |
| 1512.127463 | 2451512.631711 | 14.691 | 0.126 | 1590.121693 | 2451590.621826 | 13.838 | 0.146 |
| 1512.397993 | 2451512.902238 | 14.842 | 0.173 | 1593.104316 | 2451593.604223 | 13.675 | 0.181 |
| 1512.399003 | 2451512.903248 | 14.310 | 0.122 | 1598.107106 | 2451598.606634 | 13.591 | 0.150 |
| 1514.126423 | 2451514.630643 | 14.649 | 0.133 | 1598.107426 | 2451598.606954 | 13.603 | 0.156 |
| 1514.127433 | 2451514.631653 | 14.298 | 0.123 | 1601.127553 | 2451601.626854 | 14.480 | 0.118 |
| 1514.316653 | 2451514.820871 | 14.828 | 0.134 | 1601.128573 | 2451601.627874 | 14.643 | 0.120 |
| 1514.317673 | 2451514.821891 | 14.382 | 0.112 | 1602.128113 | 2451602.627339 | 14.500 | 0.121 |
| 1515.314613 | 2451515.818815 | 14.847 | 0.123 | 1602.129133 | 2451602.628359 | 14.330 | 0.114 |
| 1515.315623 | 2451515.819825 | 14.594 | 0.121 | 1606.130603 | 2451606.629532 | 14.516 | 0.121 |
| 1517.126573 | 2451517.630742 | 14.630 | 0.105 | 1606.131623 | 2451606.630552 | 14.288 | 0.106 |
| 1517.127583 | 2451517.631752 | 14.889 | 0.116 | 1607.131243 | 2451607.630099 | 14.578 | 0.116 |
| 1517.314903 | 2451517.819069 | 14.974 | 0.133 | 1607.132253 | 2451607.631109 | 14.336 | 0.109 |
| 1517.315923 | 2451517.820089 | 14.578 | 0.114 | 1608.131833 | 2451608.630616 | 14.677 | 0.117 |
| 1518.126593 | 2451518.630743 | 14.393 | 0.128 | 1609.132473 | 2451609.631183 | 14.531 | 0.131 |
| 1518.127613 | 2451518.631763 | 14.469 | 0.141 | 1609.133493 | 2451609.632203 | 15.000 | 0.167 |

Table 4 - Photometric data of NSVS 1925037 from the Northern Sky Variability Survey

## 3 Results

### 3.1 Analysis of photometric data from NSVS database

Initially, we analyzed with PERANSO the photometric data of Table 4 available from NSVS. The analysis of the period was performed applying three methods: ANOVA, Lomb-Scargle and CLEANest. All three methods provided consistent results among them and the following initial solution for NSVS 1925037, that does not match with ASAS-SN and WISE results of Table 3: period 29.8846 ± 1.4932 days, epoch 2451373.8545, amplitude 0.97 ± 0.25. The curve in phase for this solution is shown in Figure 4. No significative solution with a shorter period was found.

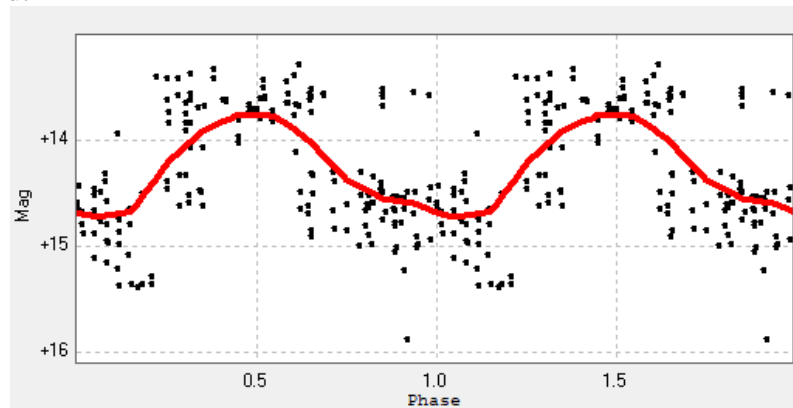

Figure 4 - Curve in phase of NSVS 1925037 (ANOVA method)

### 3.2 Analysis of photometric data from ASAS-SN

We downloaded and analyzed with PERANSO the ASAS-SN photometric data of the five stars of Table 2, searching for light curve characteristics similar to those found based on the NSVS data only or trying to identify shorter periods in accordance with the results of ASAS-SN and WISE catalogues.

For the star Gaia DR2 ID 464213469952346880, we did not find any solution with a period around 29 days, but we found a shorter period of 4.4512 ± 0.0012 days, in total agreement with the results of ASAS-SN and WISE catalogues.

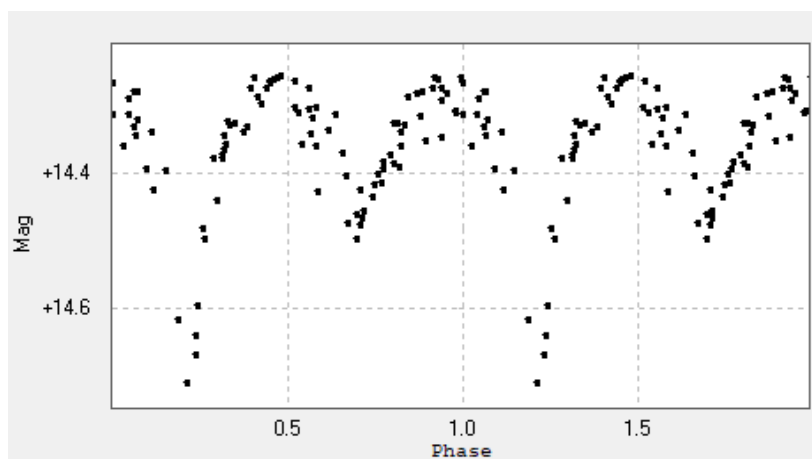

Figure 5 - Curve in phase of Gaia DR2 ID 464213469952346880

The light curve of star Gaia DR2 ID 464213469952346880 in Figure 5, we found, is typical of a β Lyrae eclipsing binary system, with a depth of the primary and secondary minimum of 0.45 and 0.21 mag. in the V filter respectively, and an epoch of the maximum of 2457037.7987 (HJD), in accordance with the ASAS-SN one.

As additional check, we analyzed the periodicity of this binary system merging the NSVS and ASAS-SN data. An offset of -0.12 mag. was added to the NSVS data to consider the difference of the two photometric systems. Figure 6 shows the result of the check: again, when NSVS data are considered only a potential 29 days period is highlighted and no significant solutions with shorter periods are found.

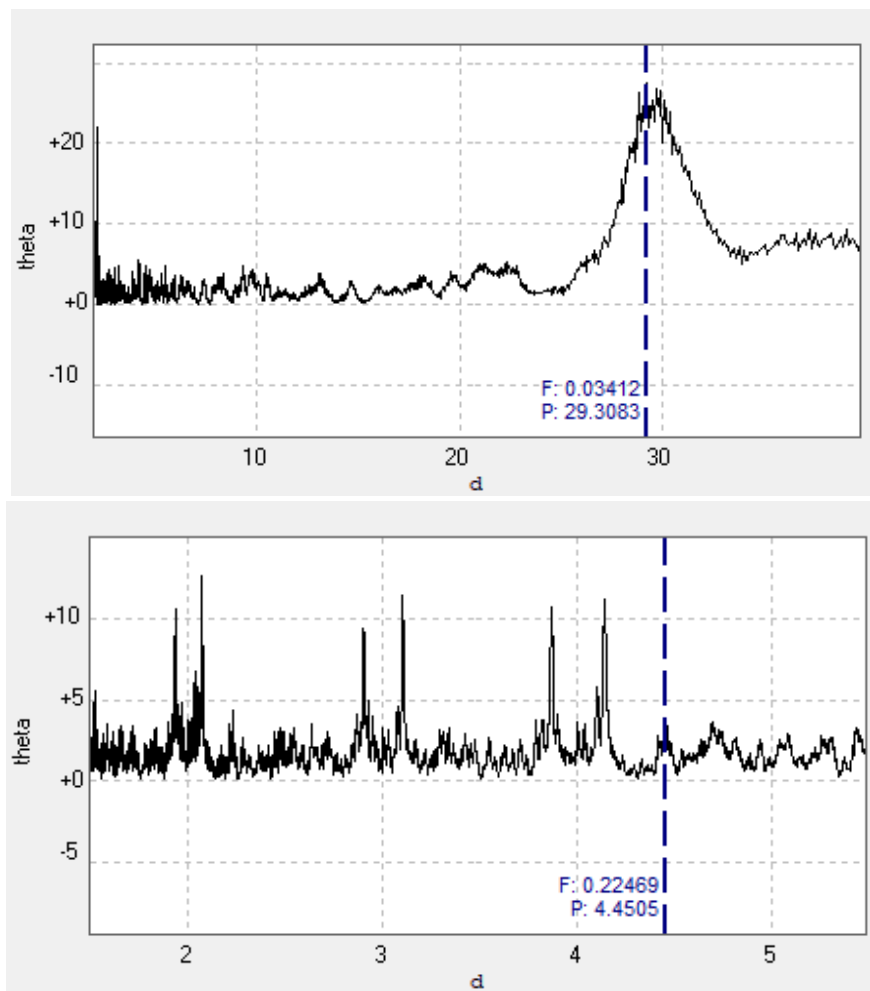

Figure 6 - Search of period with combined ASAS-SN and NSVS data (ANOVA method)

For the remaining stars in Table 2, our analysis with PERANSO did not show any variability. Figures through 7 to 10 show the curves in phase of the remaining four stars for the dominant period calculated by PERANSO, that is not reported because is deemed not significant.

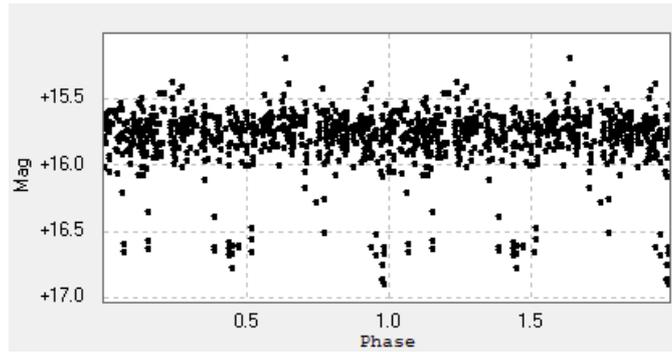

Figure 7 - Curve in phase of Gaia DR2 ID 464213504312076672

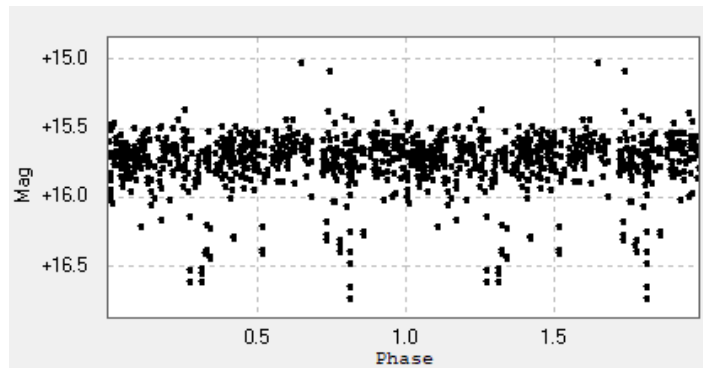

Figure 8 - Curve in phase of Gaia DR2 ID 464213710470508544

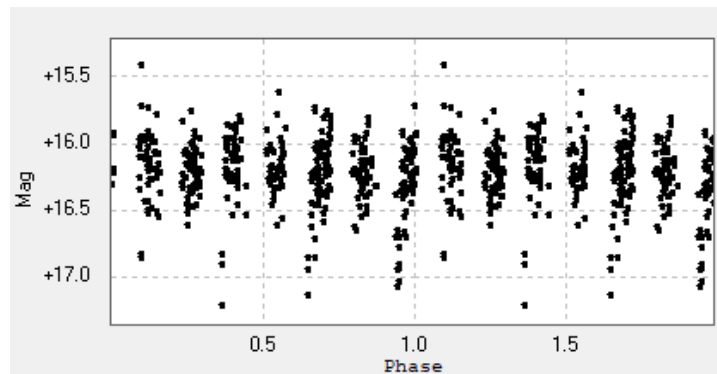

Figure 9 - Curve in phase of Gaia DR2 ID 464213504312079872

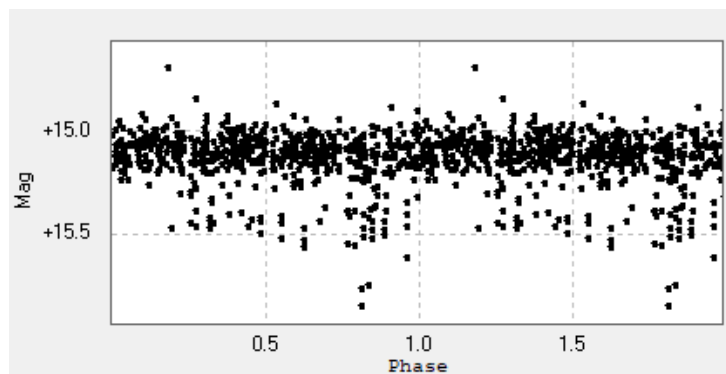

Figure 10 - Curve in phase of Gaia DR2 ID 464213676110779264

Summarizing our results, Gaia DR2 ID 464213469952346880 is a β Lyrae eclipsing binary with position and a magnitude compatible with NSVS 1925037 ones. However, the light curve shape and mostly the period found, for this variable by ASAS-SN, WISE catalogues of 4.45 days, and confirmed by our analysis, do not match with the period of about 29 days that we found analyzing only the NSVS or the combined NSVS-ASAS photometric data of NSVS 1925037.

To explain this discrepancy in the light curve shape and period we have identified the following causes:

1. NSVS 1925037 is a star other than Gaia DR2 ID 464213469952346880, this would imply that the coordinates of NSVS 1925037 are not correct, with an error greater than 60 arcsec.
2. Our analysis of the photometric NSVS measures or of the combined NSVS-ASAS ones of NSVS 1925037 is incorrect, leading to a wrong light curve shape and a ghost period of 29 days.
3. The higher spread of NSVS photometric data, greater than 2 magnitudes with respect to the ASAS-SN measures that spans in a more accurate range of 0.45 magnitudes, induces in the calculation a wrong dominant period of 29 days for the Gaia DR2 ID 464213469952346880, hiding the correct light curve and period of 4.45 days.

The attempt to identify the real cause of this discrepancy requires further investigation and will be subject of further analyses.

## 4 Conclusions

As result of our investigation we cannot establish with certainty that variable star NSVS 1925037 corresponds to any known variable or stellar object that shows variability.

ASAS-SN and WISE catalogues contain, within 10 arcsec from NSVS 1925037 coordinates, a star (ASASSN-V J023937.20+590410.9 or WISEJ023937.2+590410) that has already been reported to AAVSO as variable. This variable corresponds to star Gaia DR2 ID 464213469952346880. Our analysis of ASAS-SN photometric measurements confirms that Gaia DR2 ID 464213469952346880 is a typical β Lyrae variable, with a period of $4.4512 \pm 0.0012$ days and a depth of the primary and secondary minimum of 0.45 and 0.21 mag. in the V filter respectively.

Despite the magnitude and the position of this eclipsing binary are compatible with NSVS 1925037 data, the light curve shape and period that we found analyzing the NSVS or the combined NSVS-ASAS photometric measures are not compatible with Gaia DR2 ID 464213469952346880 characteristics.

Therefore, further analyses are required to identify a correspondence between NSVS 1925037 and a known variable or a stellar object that shows variability.

## Acknowledgements

- This activity has made use of the SIMBAD database, operated at CDS, Strasbourg, France.
- This work has made use of data from the European Space Agency (ESA) mission Gaia (https://www.cosmos.esa.int/gaia), processed by the Gaia Data Processing and Analysis Consortium (DPAC, https://www.cosmos.esa.int/web/gaia/dpac/consortium). Funding for the DPAC has been provided by national institutions, in particular the institutions participating in the Gaia Multilateral Agreement.
- This work was carried out in the context of educational and training activities provided by Italian law 'Alternanza Scuola Lavoro', July 13$^{th}$, 2015 n.107, Art.1, paragraphs 33-43.

## References

- Chen X. et al., 2018. Wide-field Infrared Survey Explorer (WISE) Catalog of Periodic Variable Stars, 2018ApJS..237...28C
- Gaia Collaboration et al. (2016): Description of the Gaia mission (spacecraft, instruments, survey and measurement principles, and operations).
- Gaia Collaboration et al. (2018b): Summary of the contents and survey properties.
- Gettel S.J., Geske M.T. & McKay T.A., 2006. A Catalog of 1022 Bright Contact Binary Stars, 2006AJ....131..621G
- Hoffman D. I., Harrison T. E. & McNamara B. J. , 2009. Automated variable star classification using the Northern Sky Variability Survey, 2009AJ....138..466H
- Kochanek C.S. et al., 2017. The All-Sky Automated Survey for Supernovae (ASAS-SN) Light Curve Server v1.0, 2017PASP..129j4502K
- Jayasinghe T. et al., (2018b). The ASAS-SN catalogue of variable stars - II. Uniform classification of 412 000 known variables, 2019MNRAS.486.1907J
- Paunzen E., Vanmunster T., 2016. Peranso - Light Curve and Period Analysis Software, arXiv:1602.05329
- Wozniak P. R. et. al., 2004. Northern Sky Variability Survey (NSVS): Public data release, arXiv:astro-ph/0401217